\documentclass{webofc}
\usepackage[varg]{txfonts}   
\usepackage{hyperref}
\usepackage{url}
\usepackage{amsmath}
\hypersetup{colorlinks=true,citecolor=blue,urlcolor=blue,linkcolor=blue}
\begin{document}
\title{Intriguing aspects of light baryon resonances}
%
%
\author{\firstname{K. P.} \lastname{Khemchandani}\inst{1}\fnsep\thanks{\email{kanchan.khemchandani@unifesp.br}} \and
        \firstname{A.} \lastname{Mart\'inez Torres}\inst{2}\fnsep\thanks{\email{amartine@if.usp.br}} \and
     \firstname{Sang-Ho} \lastname{Kim}\inst{3}    \and 
        \firstname{Seung-il} \lastname{Nam}\inst{4}     \and 
 \firstname{A.} \lastname{Hosaka}\inst{5} \and 
 \firstname{H.} \lastname{Nagahiro}\inst{6}
}

\institute{Universidade Federal de S\~ao Paulo, C.P. 01302-907, S\~ao Paulo, Brazil. 
\and
           Universidade de Sao Paulo, Instituto de Fisica, C.P. 05389-970, S\~ao Paulo,  Brazil. 
\and
Department of Physics and Origin of Matter and Evolution of Galaxies (OMEG) Institute, Soongsil University, Seoul 06978, Korea.
\and
Department of Physics, Pukyong National University (PKNU), Busan 48513, Korea Center for Extreme Nuclear Matters (CENuM).     
\and
 Research Center for Nuclear Physics (RCNP), Osaka University, Ibaraki, Osaka, 567-0047, Japan. 
 \and  Department of Physics, Nara Women’s University, Nara 630-856, Japan }

\abstract{ We discuss that some light baryon resonances exhibit properties which cannot be described when attributing a three-valence quark structure to them. Besides pointing out the hadron resonances which clearly require description beyond the quark model, we focus on the third $s_{11},~ N^*$ state and its decay to final states consisting of the lightest hyperon resonances which have a partial width comparable to that for the decay to $\pi N$. Such properties of the mentioned nucleon resonance get manifested in the cross sections and other observables related to processes producing the lightest hyperon resonances. We show that all these findings arise from the strong association of the baryon resonances to the dynamics among the ground-state hadrons. 
}
\maketitle%
\section{Introduction}
\label{intro}
Though the findings of the pentaquarks in the charm sector, as reported by the LHCb collaboration~\cite{LHCb:2015yax,LHCb:2019kea,LHCb:2020jpq} are a novelty in the field, we must recall that the concept of pentaquark nature is not new and has been discussed since the discovery of the first excited nucleon and hyperon. The puzzle of mass inversion of these first excited states, that is of $N^*(1535)$ and $\Lambda(1405)$ has been discussed in several instances~\cite{Dalitz:1959dn,Glozman:1995fu,Kaiser:1995cy,Kaiser:1996js,Oset:1997it,Inoue:2001ip,Liu:2005pm,Khemchandani:2012ki}. The enigma lies in the fact that, though the mass of the ground state $\Lambda$ is higher than the nucleons, due to the presence of the strange quark in its composition, the mass of  $\Lambda(1405)$ is lighter than that of $N^*(1535)$. The solution lies in understanding  $\Lambda(1405)$ as $uds\bar q q$, with $q=u,~d$  and  $N^*(1535)$ as $qqq\bar s s$. The same notion is frequently described as attributing a meson-baryon molecular nature to the two states. The previous statement means that $\Lambda(1405)$, for example,  arises from $\pi \Sigma$-$\bar K N$ coupled channel interaction. In fact, two complex energy poles are found in the amplitude determined by solving scattering equations when considering hadron degrees of freedom~(to see some recent reviews and well-cited works on this topic, we direct the reader to Refs.~\cite{Mai:2020ltx,Meissner:2020khl,Hyodo:2020czb,Mai:2018rjx,Roca:2017wfo,Liu:2016wxq,Kamiya:2016jqc,Molina:2015uqp,MartinezTorres:2013yma,Hall:2014uca,Menadue:2011pd,Ishii:2007ym,Takahashi:2010nj}).  The mentioned poles interfere to produce a peak on the real axis which is identified with $\Lambda(1405)$. Similarly, $N^*(1535)$ is known to couple strongly to $K\Lambda$ and $K\Sigma$ channels. 

Yet another example is that of the $\Xi(1690)$ which is the first (well-known) excited $\Xi$ state. Though it possesses a significant phase space for decay to the $\pi \Xi$ channel, its width is surprisingly small (Ref.~\cite{ParticleDataGroup:2022pth} lists $\Gamma=20\pm15$ MeV). It is known to decay mostly to $\bar K\Lambda$ and $\bar K\Sigma$ channels, which have thresholds very close to the nominal mass of $\Xi(1690)$. Within the traditional quark model,  a light quark pair would be required to be generated through the $3p_0$ model for the decay of $\Xi(1690)$ to $\pi \Xi$ as well as to $K\Lambda$. In such a case one would expect to find the partial decay width for $K\Lambda$ to be much smaller than for $\pi \Xi$, which is contradictory to the experimental findings~\cite{ParticleDataGroup:2022pth}. In Ref.~\cite{Khemchandani:2016ftn},  coupled channel interactions were studied in meson-baryon systems with total strangeness $-2$ and it was shown that $\Xi(1690)$ can be understood as a $\bar K\Sigma$ quasi-bound state. It was also found that the generated state couples very weakly to $\pi \Xi$, thus explaining the small width of $\Xi(1690)$ despite the existence of a large phase space available for such a decay process to occur.  Similar conclusions are reached by other, later works \cite{Sarti:2023wlg,Yan:2024usf,Miyahara:2016yyh,Feijoo:2023wua}.

Such findings are not only important from the spectroscopic point of view and to decipher the nature of the strong interactions at low energies but are also relevant to answering open questions in other areas of research. For instance, there exists a puzzle on the abundance of hyperons in Ar+Kcl collisions~\cite{HADES:2009mtu,HADES:2015oef}, the ratio of the yield $\Xi/(\Sigma+\Lambda)$ is found to be underestimated by the transport model. As shown in Refs.~\cite{MartinezTorres:2014son,Abreu:2016qci,Abreu:2023jcs,Abreu:2022lfy,Abreu:2022jmi}, the existence of long-lived resonances can affect the yield of hadrons in heavy ion collisions. The reason behind the former statement is that different stable hadrons produced in the collisions can interact with each other and can form a narrow state, and, in such a case,  fewer stable hadrons would reach the detectors. It was proposed in Ref.~\cite{Khemchandani:2016ftn} that, particularly in the case of the findings of Refs.~\cite{HADES:2009mtu,HADES:2015oef}, it would be important to consider that $\Xi(1690)$ can be produced by interactions of kaons with the ground state hyperons. Yet another example is the presence of an unexplained structure seen around 2 GeV, in the cross sections of the photoproduction of the $\phi$-meson on a nucleon.  In fact, it has been proposed by the authors of Ref.~\cite{Nam:2021ayk} that the structure can correspond to the excitation of a $3/2^-$ resonance whose existence was proposed in Ref.~\cite{Khemchandani:2011et}. The list of the consequences of the presence of hadron resonances can be very long. Just to mention one more example,  is the famous field of investigation of the existence of kaonic nuclear bound states with implications of the existence of strange matter in stars. This latter topic is being dedicatedly explored at the Da$\phi$ne facility in Frascati~\cite{FRASCATI-DAFNE-AMADEUS:2015wts,Curceanu:2019uph}.

It is the purpose of the present talk to review the formalism used in our works exploring the properties of light baryon resonances. As we will show, our works are based on solving scattering equations with kernels consisting of different diagrams contributing to interactions between mesons and baryons from the octets. We will focus here on a nucleon resonance found near 1.9 GeV [$N^*(1890)$], which coincides with the thresholds of kaon and light hyperons: $\Lambda(1405)$ and a proposed $\Sigma(1400)$. The existence of the latter state has been contemplated in Refs.~\cite{Oller:2000fj,Guo:2012vv,Wu:2009tu,Wu:2009nw,Gao:2010hy,Xie:2014zga,Xie:2017xwx,Khemchandani:2012ur}.  We show that the $N^*(1890)$ has a strong coupling to the kaon-light hyperon channels. Further, we discuss that the proposed nature of such a nucleon resonance is very useful in describing the cross sections of the photoproduction of $\Lambda(1405)$ near the threshold region.
 
 \section{Meson-baryon interactions}
\begin{figure}
\begin{centering}
\includegraphics[width=0.8\textwidth]{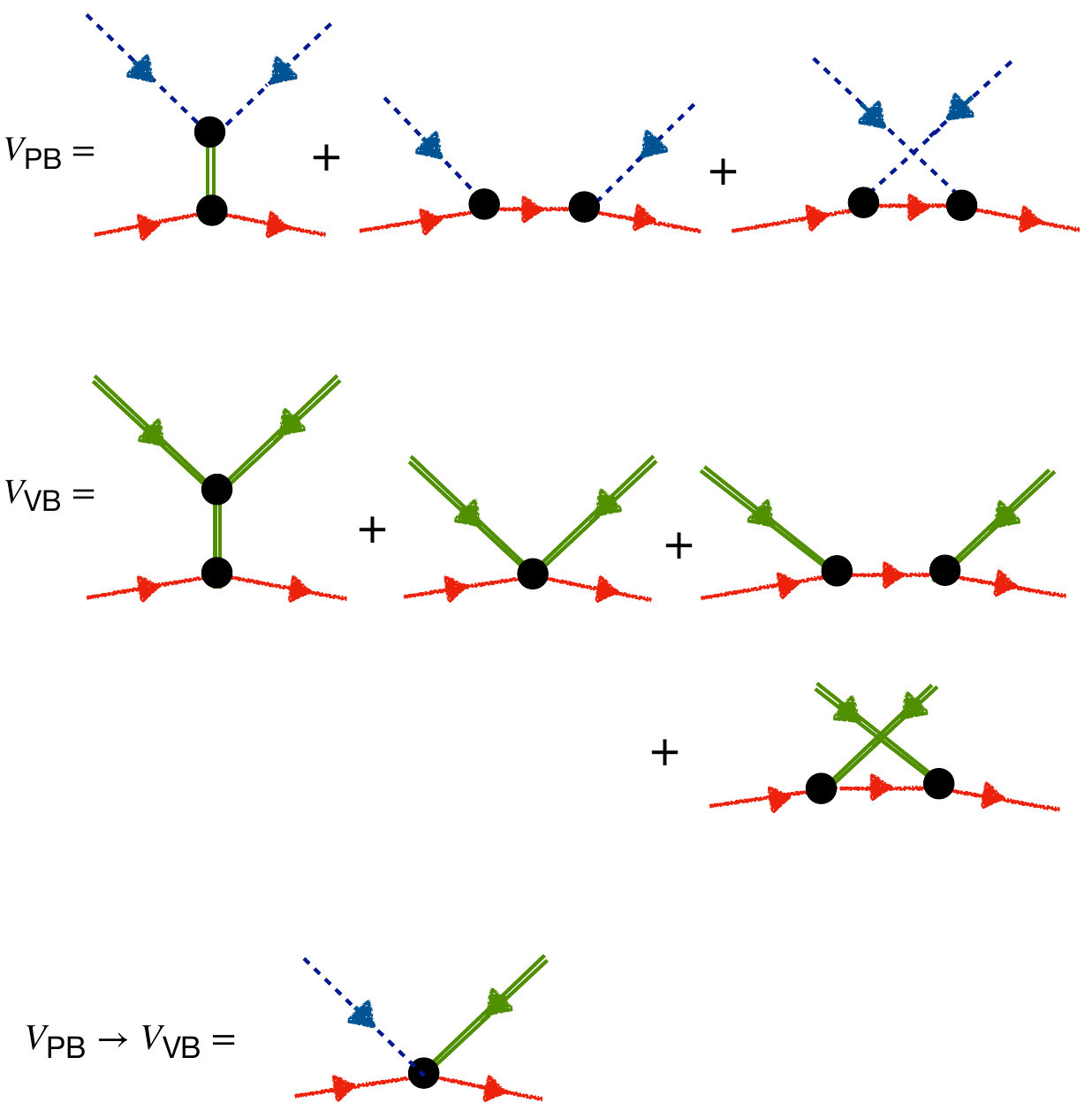}
\caption{Different diagrams considered at the tree level for meson-baryon interactions. The subscript PB (VB) stands for the pseudoscalar (vector)-baryon system. Double lines in the diagrams represent vector-mesons, dashed lines denote pseudoscalar-mesons and smeared, solid lines symbolize octet baryons. }\label{diagrams}
\end{centering}
\end{figure}

 The method which is suitable to study the formation of molecular baryon resonances, from meson-baryon interactions, is to deduce different possible interactions between the constituents of the system and then solve the Bethe-Salpeter equation,
 \begin{align}
 T=V+VGT,\label{BS}
 \end{align}
 in a coupled channel approach. The idea is to consider coupled channels made of pseudoscalars and octet-baryons, as well as vector mesons and octet-baryons. For example, to study the formation of baryon resonances with strangeness $-1$ we consider the following coupled channels: $\bar K N$, $K\Xi$, $\pi\Sigma$, $\eta \Lambda$, $\pi\Lambda$, $\eta \Sigma$, $\bar K^*N$, $K^*\Xi$, $\rho\Sigma$, $\omega \Lambda$, 
$\phi\Lambda$, $\rho \Lambda$, $\omega\Sigma$ and $\phi\Sigma$ ~\cite{Khemchandani:2018amu}. The diagrams which are considered at the tree level in Ref.~\cite{Khemchandani:2018amu} are shown in Fig.~\ref{diagrams}.
The Lagrangian used to write the different vertices shown in Fig.~\ref{diagrams} are~\cite{Khemchandani:2018amu,Khemchandani:2012ur,Khemchandani:2016ftn,Khemchandani:2012ki,Khemchandani:2013nma} 
\begin{align} \nonumber
&\mathcal{L}_{\textrm VB}= -g \Biggl\{ \langle \bar{B} \gamma_\mu \left[ V^\mu, B \right] \rangle + \langle \bar{B} \gamma_\mu B \rangle  \langle  V^\mu \rangle  
\Biggr. 
+\left. \frac{1}{4 M} \left( F \langle \bar{B} \sigma_{\mu\nu} \left[ V^{\mu\nu}, B \right] \rangle  + D \langle \bar{B} \sigma_{\mu\nu} \left\{ V^{\mu\nu}, B \right\} \rangle\right)\right\},\\\nonumber
&\mathcal{L}_{\textrm PB} = \langle \bar B i \gamma^\mu \partial_\mu B  + \bar B i \gamma^\mu[ \Gamma_\mu, B] \rangle - M_{B} \langle \bar B B \rangle
+  \frac{1}{2} D^\prime \langle \bar B \gamma^\mu \gamma_5 \{ u_\mu, B \} \rangle + \frac{1}{2} F^\prime \langle \bar B \gamma^\mu \gamma_5 [ u_\mu, B ] \rangle,\\
&\mathcal{L}_{\textrm PBVB} = \frac{-i g_{PBVB}}{2 f_\pi} \left ( F^\prime \langle \bar{B} \gamma_\mu \gamma_5 \left[ \left[ P, V_\mu \right], B \right] \rangle + 
D^\prime \langle \bar{B} \gamma_\mu \gamma_5 \left\{ \left[ P, V_\mu \right], B \right\}  \rangle \right), \label{lagrangians}
\end{align}
where $\mathcal{L}_{\textrm PB}$ is the lowest order chiral Lagrangian~\cite{Meissner:1993ah,Ecker:1994gg,Pich:1995bw,Kaiser:1995eg,Oller:2000fj,Oset:1997it,Oller:2006yh},  $\mathcal{L}_{\textrm VB}$ and $\mathcal{L}_{\textrm PBVB}$ are based on the hidden local symmetry which treats vector mesons as the gauge bosons~\cite{Bando:1984ej,Bando:1987br}.  The values of the constants $D$ = 2.4, $F$ = 0.82, $D^\prime=$0.8 and $F^\prime$=0.46, in Eq.~(\ref{lagrangians}), are taken such as to reproduce the anomalous magnetic couplings at the vector-baryon-baryon vertices in agreement with the experimental data and to get axial coupling constant of the nucleon as $F^\prime + D^\prime \simeq  g_A = 1.26$~\cite{Jenkins:1992pi,Meissner:1997hn,Jido:2002yz}. Further, $f_\pi$ represents the pion decay constant,  $g=m_v/(\sqrt{2}f_v)$ with $m_v (f_v)$ being the mass (decay constant) of the vector-meson present in the vertex, and 
\begin{align}\nonumber
\Gamma_\mu&=\frac{1}{2} \left( u^\dagger \partial_\mu u + u \partial_\mu u^\dagger  \right),\\\nonumber
 U&=u^2 = Exp \left(i \frac{P}{f_P}\right)\\\nonumber
 u_\mu&= i u^\dagger \partial_\mu U u^\dagger,
\end{align}
where the octet  fields are defined through 
\begin{eqnarray} \nonumber
&&P =
\left( \begin{array}{ccc}
\pi^0 + \frac{1}{\sqrt{3}}\eta & \sqrt{2}\pi^+ & \sqrt{2}K^{+}\\
\sqrt{2}\pi^-& -\pi^0 + \frac{1}{\sqrt{3}}\eta & \sqrt{2}K^{0}\\
\sqrt{2}K^{-} &\sqrt{2}\bar{K}^{0} & \frac{-2 }{\sqrt{3}} \eta
\end{array}\right),\\\nonumber
&&B =
\left( \begin{array}{ccc}
 \frac{1}{\sqrt{6}} \Lambda + \frac{1}{\sqrt{2}} \Sigma^0& \Sigma^+ & p\\
\Sigma^-&\frac{1}{\sqrt{6}} \Lambda- \frac{1}{\sqrt{2}} \Sigma^0 &n\\
\Xi^- &\Xi^0 & -\sqrt{\frac{2}{3}} \Lambda 
\end{array}\right).
\end{eqnarray}
The  tensor field for the vector mesons, in the Lagrangians [Eq.~(\ref{lagrangians})], is written as
\begin{equation}
V^{\mu\nu} = \partial^{\mu} V^\nu - \partial^{\nu} V^\mu + ig \left[V^\mu, V^\nu \right], \label{tensor}
\end{equation}
with, 
\begin{eqnarray}
V =\frac{1}{2}
\left( \begin{array}{ccc}
\rho^0 + \omega & \sqrt{2}\rho^+ & \sqrt{2}K^{*^+}\\
&& \\
\sqrt{2}\rho^-& -\rho^0 + \omega & \sqrt{2}K^{*^0}\\
&&\\
\sqrt{2}K^{*^-} &\sqrt{2}\bar{K}^{*^0} & \sqrt{2} \phi 
\end{array}\right)
\end{eqnarray}
in our normalization scheme. It is worth mentioning that the two-meson field part of the Eq.~(\ref{tensor}) leads to a contact term when substituting it in Eq.~(\ref{lagrangians}), and hence giving rise to an extra diagram contributing to vector-baryon interaction when compared to the pseudoscalar-baryon systems.

\section{Amplitudes showing appearance of resonances}
\begin{figure}[h!]
\includegraphics[width=0.45\textwidth]{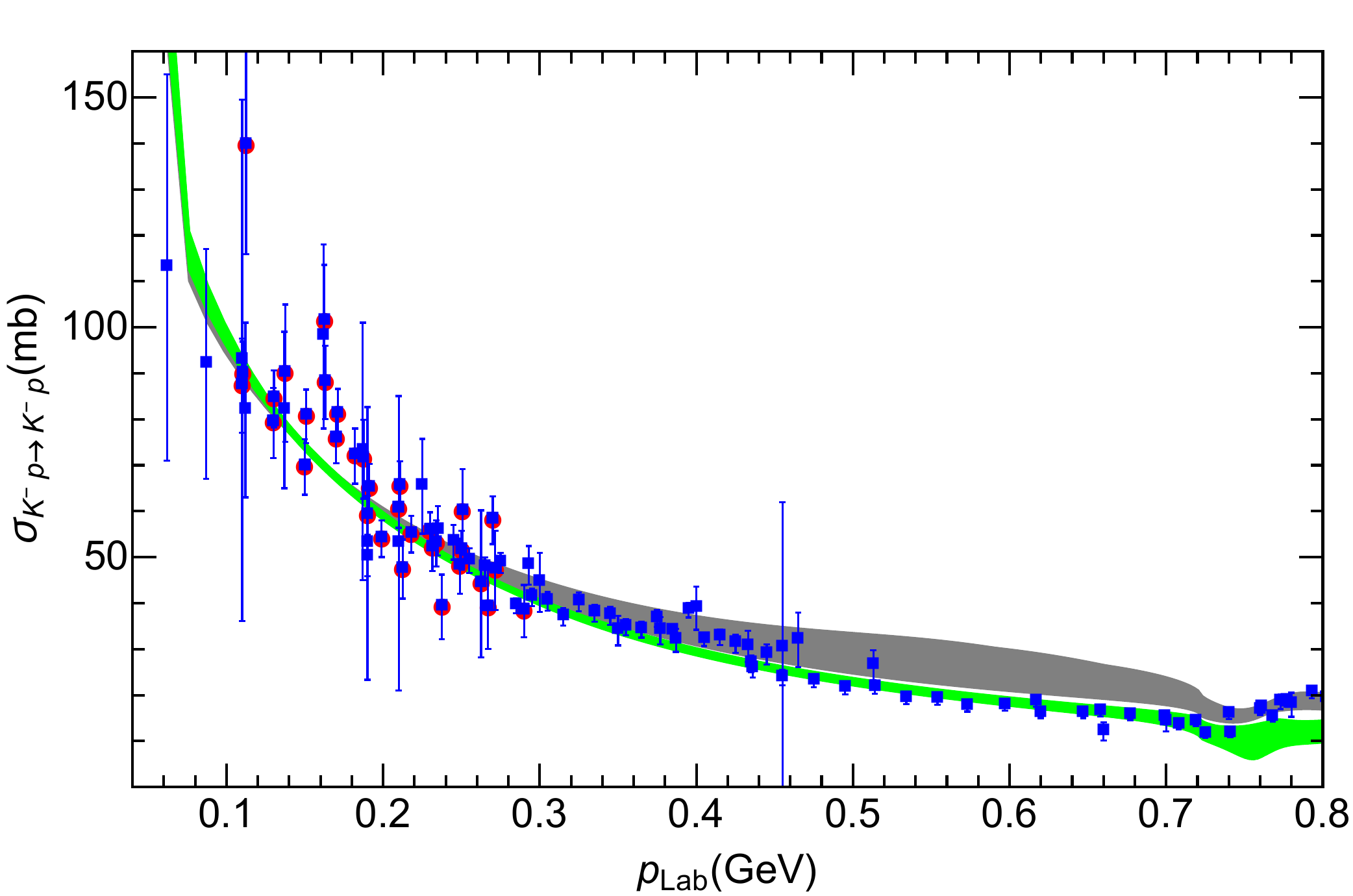}\includegraphics[width=0.45\textwidth]{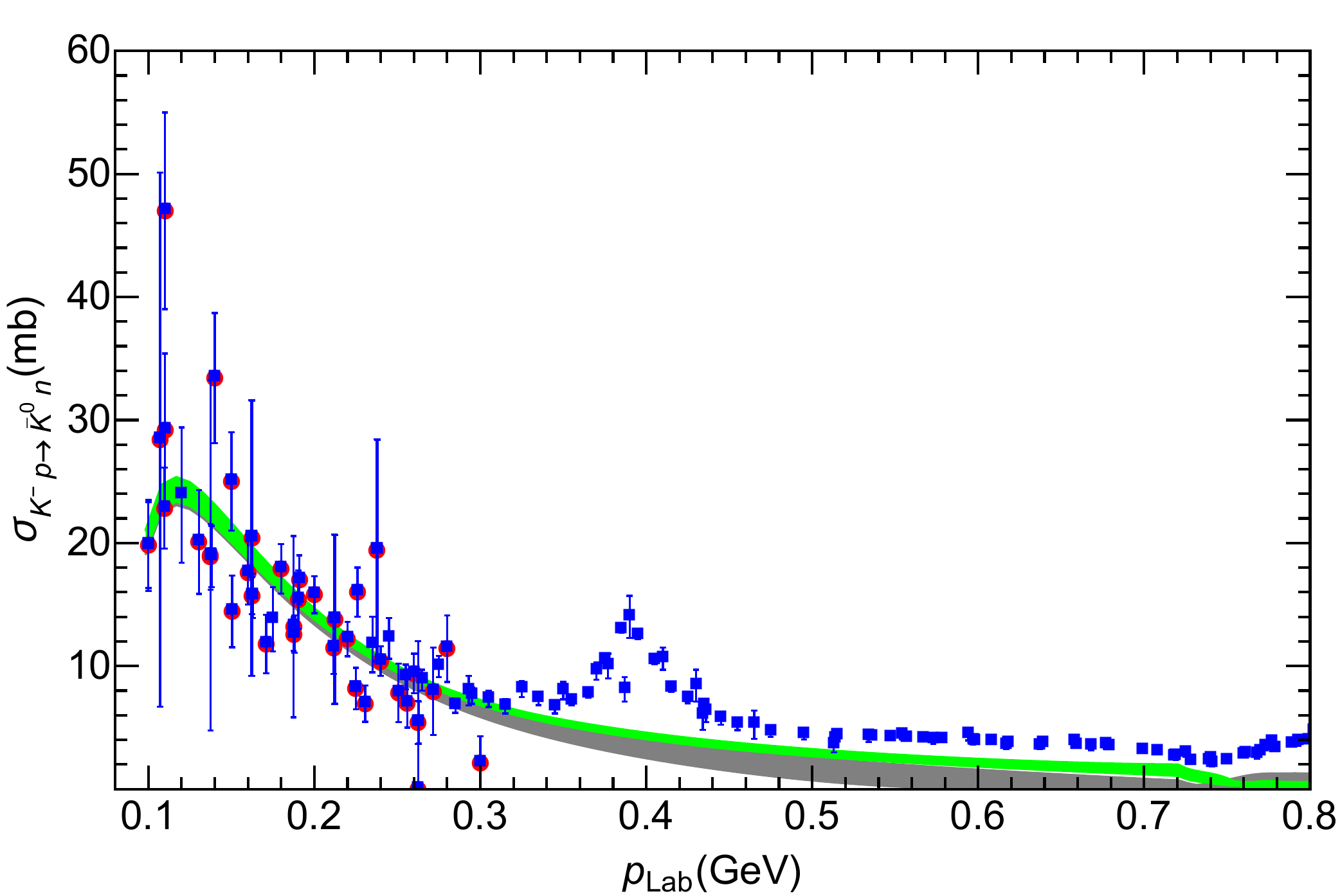}
\caption{Total cross sections for the $K^- p \to K^- p$ (left panel) and $K^- p \to \bar K^0 n$ (right panel) processes.  The shaded regions show two types of similar quality fits to the experimental data. The  data are taken from Refs.~\cite{Humphrey:1962zz,Ciborowski:1982et,Evans:1983hz,Kim:1967zze,Sakitt:1965kh,Kittel:1966zz}. }\label{xnKbarN}
\end{figure}

Having determined the tree-level diagrams, Eq.~(\ref{BS}) is solved in the on-shell approximation, in which case we end up with a divergent loop function, that needs to be regularized. The regularization parameters, which are subtraction constants related to the diagonal elements of the $T$-matrix, together with $g_\pi$ and $g_{PBVB}$ are taken as free parameters whose values are fixed by making a $\chi^2$-fit to relevant experimental data. For instance, in the case of meson-baryon systems with total strangeness $-1$, we consider~\cite{Khemchandani:2018amu} experimental data on the total cross sections of the kaon-nucleon processes: $K^- p \to K^- p,~\bar K^0 n,~ \eta \Lambda,~\pi^0 \Lambda~\pi^0 \Sigma^0,~ \pi^\pm \Sigma^\mp$, near the related threshold region ( about 30-50 MeV above the respective thresholds), as well as the data on energy level shift and width of the $1s$-state of the kaonic hydrogen determined by the SIDDHARTA collaboration~\cite{SIDDHARTA:2011dsy}. In the case of systems with total strangeness zero, we consider the data on isospin 1/2 and  3/2 $\pi N$ amplitudes, available from the partial wave analysis of Ref.~\cite{Arndt:1995bj}, and cross sections on the $\pi^- p \to \eta n$  and $\pi^- p \to K^0 \Lambda$ processes. As an example, we show two types of fits obtained for the total cross section for $K^- p \to K^- p$ and $K^- p \to \bar K^0 n$ processes in Fig.~\ref{xnKbarN}. More results can be found in Refs.~\cite{Khemchandani:2018amu,Khemchandani:2013nma}.

Before further discussions, we must recall that the motivation of the formalism discussed here is to study the contribution of hadron dynamics to the structure of resonances. When poles in the complex plane are found to appear in the amplitudes determined by solving the Bethe-Salpeter equation, the related states are interpreted as those that require contributions beyond the one attributed by the traditional quark model.  For such purposes, all the interactions are kept in s-wave. Thus, vector-baryon systems with a particular strangeness can have total spin-parity $1/2^-$ or $3/2^-$.  While studying states with spin-parity $1/2^-$,  systems made of both pseudoscalar and vector mesons can be treated as coupled channels. Such a treatment is especially important when the two types of channels can have similar thresholds, for example, $K\Xi$ and $\bar K^* N$. 
Another advantage of the mentioned formalism is the possibility of determining the coupling of the low-lying resonances, like $\Lambda(1405)$ to the vector-baryon channels. Such information can be used to determine the radiative decay of baryon resonances, through the vector meson dominance, which in turn can serve for studying the photoproduction of resonances. 

\begin{figure}[h!]
\centering
\includegraphics[width=0.35\textwidth]{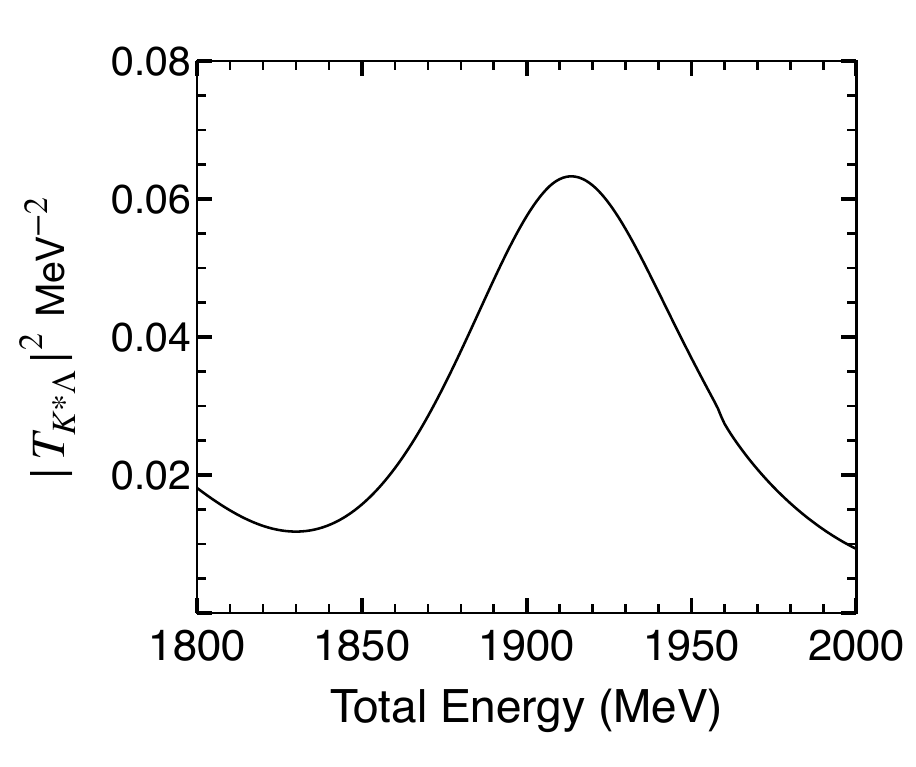}
\caption{Modulus squared amplitude for $ K^* \Lambda\to  K^*\Lambda$.  }\label{KstrLambda}
\end{figure}

We focus in this proceedings, especially on reviewing the results of the two particular cases: of systems with total strangeness zero and strangeness $-1$ having total spin-parity $1/2^-$. As we will discuss, they are of special interest since the results of the two cases can be interrelated and a compound effect of their nature can be observed in data.   In the case of strangeness 0, we find the appearance of three $S_{11}$ nucleon resonances: $N^*(1535)$, $N^*(1650)$, and $N^*(1895)$ in the isospin 1/2 sector and of $\Delta(1620)$ in total isospin 3/2. It is important to mention that $N^*(1535)$ is found to strongly couple to the $K\Sigma$ channel, in agreement with the findings of Refs.~\cite{Kaiser:1995cy,Kaiser:1996js,Inoue:2001ip,Liu:2005pm}, while $N^*(1650)$ couples most to  $\rho N$ (in agreement with the results of Refs.~\cite{Garzon:2012np,Garzon:2014ida}). The case of $N^*(1895)$ is also interesting since the traditional quark model predicts the existence of a  third $S_{11}$ state to be around 2 GeV~\cite{Isgur:1978xj,Bijker:1994yr,Hosaka:1997kh,Takayama:1999kc}, indicating the need of considering contributions from hadron dynamics to describe its properties (as indeed found in Ref.~\cite{Khemchandani:2013nma}). In Fig.~\ref{KstrLambda} we show the modulus squared amplitude for the $ K^* \Lambda$ channel, which shows a clear peak related to $N^*(1895)$.

In the systems with strangeness $-1$, we find that poles related to $\Lambda(1405)$, $\Lambda(1670)$ and $\Lambda(1800)$ arise in isoscalar amplitudes as well as the isovector states which can be related to $\Sigma(1400)$, to either $\Sigma(1620)$ or $\Sigma(1670)$ and $\Sigma(1900)$ are found to appear. 
It must be mentioned that, out of the aforementioned states,  $\Sigma(1400)$ is not a well-established state.  There exist several works indicating that different experimental data require the existence of $\Sigma(1400)$ to get a good fit~\cite{Oller:2000fj,Guo:2012vv,Wu:2009tu,Wu:2009nw,Gao:2010hy,Xie:2014zga,Xie:2017xwx,Khemchandani:2012ur}. Our work, in Ref.~\cite{Khemchandani:2018amu}, also led to such conclusions. However, better quality data are required to make stronger affirmations. As we shall shortly discuss, data on photon-proton collisions can be useful to clarify the situation.

\section{Implications of the properties of $N^*(1895)$ on the photoproduction of lowest hyperon resonances}
\begin{figure}[h!]
\centering
\includegraphics[width=0.5\textwidth]{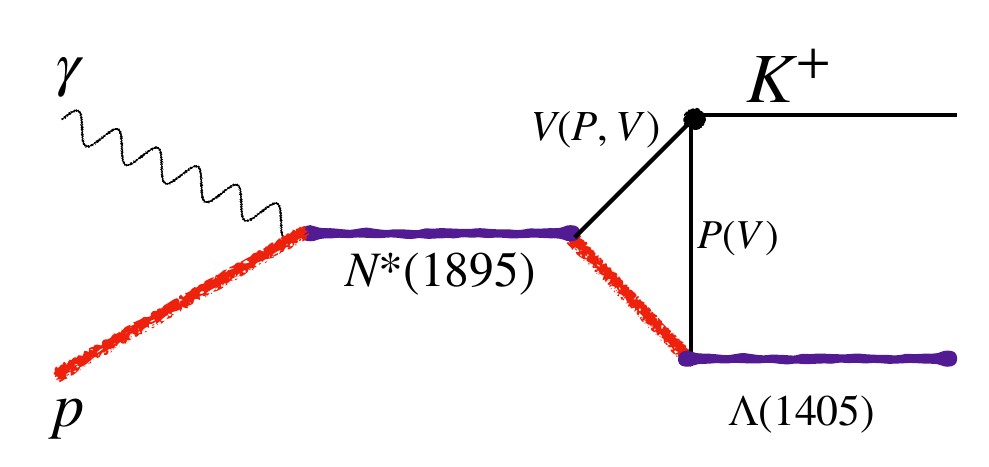}
\caption{A diagram showing exchange of a nucleon resonance in the $s$-channel, leading to the production of a hyperon resonance in the final state.}\label{diagram2}
\end{figure}

At this point, we can present a study of the photoproduction process as a tool to investigate the properties of nucleon as well as hyperon resonances. Specifically interrelated  states are $N^*(1895)$, $\Lambda(1405)$ and the proposed $\Sigma(1400)$ since the mass of $N^*(1895)$ lies very close to the thresholds of kaon and hyperon(1400). In such a case a diagram, as shown in Fig.~\ref{diagram2}, can contribute to the photoproduction of $\Lambda(1405)$ and $\Sigma(1400)$.
It is important to mention here that all the vertices shown in Fig.~\ref{diagram2} are well determined in our works and are not free parameters. The radiative decay of $N^*(1895)$ can be straightforwardly calculated through vector meson dominance since the couplings of the resonance to vector-baryon channels are determined as residues of the related poles found in the complex energy plane. Similarly, all the vertices shown in the triangular loop are known as couplings of either $N^*(1895)$ or of $\Lambda(1405)$ to different meson-baryon channels. More details of the calculations can be found in Refs.~\cite{Khemchandani:2020exc,Kim:2021wov,Khemchandani:2022yyx,Khemchandani:2022dst}.

With this idea in mind, in Ref.~\cite{Kim:2021wov}, we first investigated $\gamma+p\to K^+\Lambda(1405)$ by considering an exchange of kaon and $K^*(892)$ Regge series in the $t$-channel, the exchange of a nucleon and several resonances in the $s$-channel and those of hyperons in the $u$-channel diagram. We compared our results with the experimental data available on the process from the CLAS Collaboration~\cite{CLAS:2013rjt,CLAS:2013rxx} and found that the total cross sections near the threshold get a large contribution from the $s$-channel exchange of $N^*(1895)$. Having results in good comparison with the data, we then studied $\gamma+p\to K^+\Sigma(1400)$ and found that the total cross sections for this process are about an order magnitude smaller than the ones for the photoproduction of $\Lambda(1405)$ but can be measured at experimental facilities like JLAB, LEPS, etc. We have provided results for several polarization variables too,  in Ref.~\cite{Kim:2021wov}, for the photoproduction of both $\Lambda(1405)$ and $\Sigma(1400)$, which we hope can motivate experimental investigations of the photoproduction processes in the future. 

Another takeaway message of our works is that the partial wave analyses of the data on photoproduction of the low-lying hyperon resonances can be used for extracting information on nucleon resonances around 1900 MeV.  

\section{Acknowledgements}

K.P.K and A.M.T  thank the partial support provided by the Brazilian funding agencies CNPq ( Grant numbers 306461/2023-4, 304510/2023-8 and 407437/2023-1 ), FAPESP ( Grant Numbers 2022/08347-9, and  2023/01182-7) as well as INCT-F\'{\i}sica Nuclear e Aplicações (Contract No. 464898/2014-5). This work is also partly supported by the National Research Foundation (NRF) grants funded by the Korean government (MSIT) (Nos. 2018R1A5A1025563, 2022R1A2C1003964, and 2022K2A9A1A0609176).

\bibliography{SNrefs}

\begin{thebibliography}{81}

\bibitem{LHCb:2015yax}
R.~Aaij et~al. (LHCb), Phys. Rev. Lett. \textbf{115}, 072001 (2015),
  \texttt{1507.03414}

\bibitem{LHCb:2019kea}
R.~Aaij et~al. (LHCb), Phys. Rev. Lett. \textbf{122}, 222001 (2019),
  \texttt{1904.03947}

\bibitem{LHCb:2020jpq}
R.~Aaij et~al. (LHCb), Sci. Bull. \textbf{66}, 1278 (2021), \texttt{2012.10380}

\bibitem{Dalitz:1959dn}
R.H. Dalitz, S.F. Tuan, Phys. Rev. Lett. \textbf{2}, 425 (1959)

\bibitem{Glozman:1995fu}
L.Y. Glozman, D.O. Riska, Phys. Rept. \textbf{268}, 263 (1996),
  \texttt{hep-ph/9505422}

\bibitem{Kaiser:1995cy}
N.~Kaiser, P.B. Siegel, W.~Weise, Phys. Lett. B \textbf{362}, 23 (1995),
  \texttt{nucl-th/9507036}

\bibitem{Kaiser:1996js}
N.~Kaiser, T.~Waas, W.~Weise, Nucl. Phys. A \textbf{612}, 297 (1997),
  \texttt{hep-ph/9607459}

\bibitem{Oset:1997it}
E.~Oset, A.~Ramos, Nucl. Phys. A \textbf{635}, 99 (1998),
  \texttt{nucl-th/9711022}

\bibitem{Inoue:2001ip}
T.~Inoue, E.~Oset, M.J. Vicente~Vacas, Phys. Rev. C \textbf{65}, 035204 (2002),
  \texttt{hep-ph/0110333}

\bibitem{Liu:2005pm}
B.C. Liu, B.S. Zou, Phys. Rev. Lett. \textbf{96}, 042002 (2006),
  \texttt{nucl-th/0503069}

\bibitem{Khemchandani:2012ki}
K.P. Khemchandani, A.~Martinez~Torres, H.~Kaneko, H.~Nagahiro, A.~Hosaka, J.
  Phys. Conf. Ser. \textbf{374}, 012007 (2012), \texttt{1202.3840}

\bibitem{Mai:2020ltx}
M.~Mai, Eur. Phys. J. ST \textbf{230}, 1593 (2021), \texttt{2010.00056}

\bibitem{Meissner:2020khl}
U.G. Mei\ss{}ner, Symmetry \textbf{12}, 981 (2020), \texttt{2005.06909}

\bibitem{Hyodo:2020czb}
T.~Hyodo, M.~Niiyama, Prog. Part. Nucl. Phys. \textbf{120}, 103868 (2021),
  \texttt{2010.07592}

\bibitem{Mai:2018rjx}
M.~Mai, Few Body Syst. \textbf{59}, 61 (2018)

\bibitem{Roca:2017wfo}
L.~Roca, J.~Nieves, E.~Oset, JPS Conf. Proc. \textbf{17}, 071002 (2017)

\bibitem{Liu:2016wxq}
Z.W. Liu, J.M.M. Hall, D.B. Leinweber, A.W. Thomas, J.J. Wu, Phys. Rev. D
  \textbf{95}, 014506 (2017), \texttt{1607.05856}

\bibitem{Kamiya:2016jqc}
Y.~Kamiya, K.~Miyahara, S.~Ohnishi, Y.~Ikeda, T.~Hyodo, E.~Oset, W.~Weise,
  Nucl. Phys. A \textbf{954}, 41 (2016), \texttt{1602.08852}

\bibitem{Molina:2015uqp}
R.~Molina, M.~D\"oring, Phys. Rev. D \textbf{94}, 056010 (2016), [Addendum:
  Phys.Rev.D 94, 079901 (2016)], \texttt{1512.05831}

\bibitem{MartinezTorres:2013yma}
A.~Martinez~Torres, M.~Bayar, D.~Jido, E.~Oset, Int. J. Mod. Phys. Conf. Ser.
  \textbf{26}, 1460057 (2014), \texttt{1309.6681}

\bibitem{Hall:2014uca}
J.M.M. Hall, W.~Kamleh, D.B. Leinweber, B.J. Menadue, B.J. Owen, A.W. Thomas,
  R.D. Young, Phys. Rev. Lett. \textbf{114}, 132002 (2015), \texttt{1411.3402}

\bibitem{Menadue:2011pd}
B.J. Menadue, W.~Kamleh, D.B. Leinweber, M.S. Mahbub, Phys. Rev. Lett.
  \textbf{108}, 112001 (2012), \texttt{1109.6716}

\bibitem{Ishii:2007ym}
N.~Ishii, T.~Doi, M.~Oka, H.~Suganuma, Prog. Theor. Phys. Suppl. \textbf{168},
  598 (2007), \texttt{0707.0079}

\bibitem{Takahashi:2010nj}
T.T. Takahashi, M.~Oka, Prog. Theor. Phys. Suppl. \textbf{186}, 172 (2010),
  \texttt{1009.1790}

\bibitem{ParticleDataGroup:2022pth}
R.L. Workman et~al. (Particle Data Group), PTEP \textbf{2022}, 083C01 (2022)

\bibitem{Khemchandani:2016ftn}
K.P. Khemchandani, A.~Mart\'\i{}nez~Torres, A.~Hosaka, H.~Nagahiro, F.S.
  Navarra, M.~Nielsen, Phys. Rev. D \textbf{97}, 034005 (2018),
  \texttt{1608.07086}

\bibitem{Sarti:2023wlg}
V.M. Sarti, A.~Feijoo, I.~Vida\~na, A.~Ramos, F.~Giacosa, T.~Hyodo, Y.~Kamiya
  (2023), \texttt{2309.08756}

\bibitem{Yan:2024usf}
Y.~Yan, Q.~Huang, X.~Zhu, H.~Huang, J.~Ping (2024), \texttt{2404.14753}

\bibitem{Miyahara:2016yyh}
K.~Miyahara, T.~Hyodo, M.~Oka, J.~Nieves, E.~Oset, Phys. Rev. C \textbf{95},
  035212 (2017), \texttt{1609.00895}

\bibitem{Feijoo:2023wua}
A.~Feijoo, V.~Valcarce~Cadenas, V.K. Magas, Phys. Lett. B \textbf{841}, 137927
  (2023), \texttt{2303.01323}

\bibitem{HADES:2009mtu}
G.~Agakishiev et~al. (HADES), Phys. Rev. Lett. \textbf{103}, 132301 (2009),
  \texttt{0907.3582}

\bibitem{HADES:2015oef}
G.~Agakishiev et~al. (HADES), Eur. Phys. J. A \textbf{52}, 178 (2016),
  \texttt{1512.07070}

\bibitem{MartinezTorres:2014son}
A.~Martinez~Torres, K.P. Khemchandani, F.S. Navarra, M.~Nielsen, L.M. Abreu,
  Phys. Rev. D \textbf{90}, 114023 (2014), [Erratum: Phys.Rev.D 93, 059902
  (2016)], \texttt{1405.7583}

\bibitem{Abreu:2016qci}
L.M. Abreu, K.P. Khemchandani, A.~Martinez~Torres, F.S. Navarra, M.~Nielsen,
  Phys. Lett. B \textbf{761}, 303 (2016), \texttt{1604.07716}

\bibitem{Abreu:2023jcs}
L.M. Abreu, A.L.M. Britto, F.S. Navarra, H.P.L. Vieira, Phys. Rev. D
  \textbf{109}, 014041 (2024), \texttt{2310.03948}

\bibitem{Abreu:2022lfy}
L.M. Abreu, H.P.L. Vieira, F.S. Navarra, Phys. Rev. D \textbf{105}, 116029
  (2022), \texttt{2202.10882}

\bibitem{Abreu:2022jmi}
L.M. Abreu, F.S. Navarra, H.P.L. Vieira, Phys. Rev. D \textbf{106}, 076001
  (2022), \texttt{2206.03399}

\bibitem{Nam:2021ayk}
S.I. Nam, Phys. Rev. D \textbf{103}, 054040 (2021), \texttt{2101.03317}

\bibitem{Khemchandani:2011et}
K.P. Khemchandani, H.~Kaneko, H.~Nagahiro, A.~Hosaka, Phys. Rev. D \textbf{83},
  114041 (2011), \texttt{1104.0307}

\bibitem{FRASCATI-DAFNE-AMADEUS:2015wts}
A.~Scordo et~al. (FRASCATI-DAFNE-AMADEUS), AIP Conf. Proc. \textbf{1735},
  080015 (2016), \texttt{1512.06555}

\bibitem{Curceanu:2019uph}
C.~Curceanu et~al., Rev. Mod. Phys. \textbf{91}, 025006 (2019)

\bibitem{Oller:2000fj}
J.A. Oller, U.G. Meissner, Phys. Lett. B \textbf{500}, 263 (2001),
  \texttt{hep-ph/0011146}

\bibitem{Guo:2012vv}
Z.H. Guo, J.A. Oller, Phys. Rev. C \textbf{87}, 035202 (2013),
  \texttt{1210.3485}

\bibitem{Wu:2009tu}
J.J. Wu, S.~Dulat, B.S. Zou, Phys. Rev. D \textbf{80}, 017503 (2009),
  \texttt{0906.3950}

\bibitem{Wu:2009nw}
J.J. Wu, S.~Dulat, B.S. Zou, Phys. Rev. C \textbf{81}, 045210 (2010),
  \texttt{0909.1380}

\bibitem{Gao:2010hy}
P.~Gao, J.J. Wu, B.S. Zou, Phys. Rev. C \textbf{81}, 055203 (2010),
  \texttt{1001.0805}

\bibitem{Xie:2014zga}
J.J. Xie, J.J. Wu, B.S. Zou, Phys. Rev. C \textbf{90}, 055204 (2014),
  \texttt{1407.7984}

\bibitem{Xie:2017xwx}
J.J. Xie, L.S. Geng, Phys. Rev. D \textbf{95}, 074024 (2017),
  \texttt{1703.09502}

\bibitem{Khemchandani:2012ur}
K.P. Khemchandani, A.~Martinez~Torres, H.~Nagahiro, A.~Hosaka, Phys. Rev. D
  \textbf{85}, 114020 (2012), \texttt{1203.6711}

\bibitem{Khemchandani:2018amu}
K.P. Khemchandani, A.~Mart\'\i{}nez~Torres, J.A. Oller, Phys. Rev. C
  \textbf{100}, 015208 (2019), \texttt{1810.09990}

\bibitem{Khemchandani:2013nma}
K.P. Khemchandani, A.~Martinez~Torres, H.~Nagahiro, A.~Hosaka, Phys. Rev. D
  \textbf{88}, 114016 (2013), \texttt{1307.8420}

\bibitem{Meissner:1993ah}
U.G. Meissner, Rept. Prog. Phys. \textbf{56}, 903 (1993),
  \texttt{hep-ph/9302247}

\bibitem{Ecker:1994gg}
G.~Ecker, Prog. Part. Nucl. Phys. \textbf{35}, 1 (1995),
  \texttt{hep-ph/9501357}

\bibitem{Pich:1995bw}
A.~Pich, Rept. Prog. Phys. \textbf{58}, 563 (1995), \texttt{hep-ph/9502366}

\bibitem{Kaiser:1995eg}
N.~Kaiser, P.B. Siegel, W.~Weise, Nucl. Phys. A \textbf{594}, 325 (1995),
  \texttt{nucl-th/9505043}

\bibitem{Oller:2006yh}
J.A. Oller, M.~Verbeni, J.~Prades, JHEP \textbf{09}, 079 (2006),
  \texttt{hep-ph/0608204}

\bibitem{Bando:1984ej}
M.~Bando, T.~Kugo, S.~Uehara, K.~Yamawaki, T.~Yanagida, Phys. Rev. Lett.
  \textbf{54}, 1215 (1985)

\bibitem{Bando:1987br}
M.~Bando, T.~Kugo, K.~Yamawaki, Phys. Rept. \textbf{164}, 217 (1988)

\bibitem{Jenkins:1992pi}
E.E. Jenkins, M.E. Luke, A.V. Manohar, M.J. Savage, Phys. Lett. B \textbf{302},
  482 (1993), [Erratum: Phys.Lett.B 388, 866--866 (1996)],
  \texttt{hep-ph/9212226}

\bibitem{Meissner:1997hn}
U.G. Meissner, S.~Steininger, Nucl. Phys. B \textbf{499}, 349 (1997),
  \texttt{hep-ph/9701260}

\bibitem{Jido:2002yz}
D.~Jido, A.~Hosaka, J.C. Nacher, E.~Oset, A.~Ramos, Phys. Rev. C \textbf{66},
  025203 (2002), \texttt{hep-ph/0203248}

\bibitem{SIDDHARTA:2011dsy}
M.~Bazzi et~al. (SIDDHARTA), Phys. Lett. B \textbf{704}, 113 (2011),
  \texttt{1105.3090}

\bibitem{Arndt:1995bj}
R.A. Arndt, I.I. Strakovsky, R.L. Workman, M.M. Pavan, Phys. Rev. C
  \textbf{52}, 2120 (1995), \texttt{nucl-th/9505040}

\bibitem{Humphrey:1962zz}
W.E. Humphrey, R.R. Ross, Phys. Rev. \textbf{127}, 1305 (1962)

\bibitem{Ciborowski:1982et}
J.~Ciborowski et~al., J. Phys. G \textbf{8}, 13 (1982)

\bibitem{Evans:1983hz}
D.~Evans, J.V. Major, E.~Rondio, J.A. Zakrzewski, J.E. Conboy, D.J. Miller,
  T.~Tymieniecka, J. Phys. G \textbf{9}, 885 (1983)

\bibitem{Kim:1967zze}
J.K. Kim, Phys. Rev. Lett. \textbf{19}, 1074 (1967)

\bibitem{Sakitt:1965kh}
M.~Sakitt, T.B. Day, R.G. Glasser, N.~Seeman, J.H. Friedman, W.E. Humphrey,
  R.R. Ross, Phys. Rev. \textbf{139}, B719 (1965)

\bibitem{Kittel:1966zz}
W.~Kittel, G.~Otter, I.~Wacek, Phys. Lett. \textbf{21}, 349 (1966)

\bibitem{Garzon:2012np}
E.J. Garzon, E.~Oset, Eur. Phys. J. A \textbf{48}, 5 (2012), \texttt{1201.3756}

\bibitem{Garzon:2014ida}
E.J. Garzon, E.~Oset, Phys. Rev. C \textbf{91}, 025201 (2015),
  \texttt{1411.3547}

\bibitem{Isgur:1978xj}
N.~Isgur, G.~Karl, Phys. Rev. D \textbf{18}, 4187 (1978)

\bibitem{Bijker:1994yr}
R.~Bijker, F.~Iachello, A.~Leviatan, Annals Phys. \textbf{236}, 69 (1994),
  \texttt{nucl-th/9402012}

\bibitem{Hosaka:1997kh}
A.~Hosaka, H.~Toki, M.~Takayama, Mod. Phys. Lett. A \textbf{13}, 1699 (1998),
  \texttt{hep-ph/9711295}

\bibitem{Takayama:1999kc}
M.~Takayama, H.~Toki, A.~Hosaka, Prog. Theor. Phys. \textbf{101}, 1271 (1999)

\bibitem{Khemchandani:2020exc}
K.P. Khemchandani, A.~Martinez~Torres, H.~Nagahiro, A.~Hosaka, Phys. Rev. D
  \textbf{103}, 016015 (2021), \texttt{2010.04584}

\bibitem{Kim:2021wov}
S.H. Kim, K.P. Khemchandani, A.~Martinez~Torres, S.i. Nam, A.~Hosaka, Phys.
  Rev. D \textbf{103}, 114017 (2021), \texttt{2101.08668}

\bibitem{Khemchandani:2022yyx}
K.P. Khemchandani, A.~Mart\'\i{}nez~Torres, S.H. Kim, S.i. Nam, H.~Nagahiro,
  A.~Hosaka, Rev. Mex. Fis. Suppl. \textbf{3}, 0308063 (2022),
  \texttt{2201.06471}

\bibitem{Khemchandani:2022dst}
K.P. Khemchandani, A.~Martinez~Torres, S.H. Kim, S.i. Nam, A.~Hosaka, Acta
  Phys. Polon. A \textbf{142}, 329 (2022), \texttt{2211.14167}

\bibitem{CLAS:2013rjt}
K.~Moriya et~al. (CLAS), Phys. Rev. C \textbf{87}, 035206 (2013),
  \texttt{1301.5000}

\bibitem{CLAS:2013rxx}
K.~Moriya et~al. (CLAS), Phys. Rev. C \textbf{88}, 045201 (2013), [Addendum:
  Phys.Rev.C 88, 049902 (2013)], \texttt{1305.6776}

\end{thebibliography}

\end{document}